\documentclass[HARVARD,Times2COL]{WileyNJDv5} %STIX1COL,STIX2COL,STIXSMALL
\usepackage{soul, xcolor}

\articletype{Data Services Article}%

\received{Date Month Year}
\revised{Date Month Year}
\accepted{Date Month Year}
\journal{Journal}
\volume{00}
\copyyear{2023}
\startpage{1}

\begin{document}

\title{The Sand Atlas}

\author[1]{Ilija Vego}
\author[1,*]{Benjy Marks}

\authormark{VEGO \textsc{et al.}}
\titlemark{THE SAND ATLAS}

\address[1]{\orgdiv{School of Civil Engineering}, \orgname{The University of Sydney}, \orgaddress{\state{NSW}, \country{Australia}}}

\corres{Corresponding author Benjy Marks, School of Civil Engineering, The University of Sydney, NSW, Australia, 2006\email{benjy.marks@sydney.edu.au}}

% \presentaddress{This is sample for present address text this is sample for present address text.}

%\fundingInfo{Text}
%\JELinfo{ejlje}

\abstract[Abstract]{The Sand Atlas is a publicly accessible repository dedicated to the collection, processing, and sharing of high-resolution 3D models of sand-sized particles. This dataset offers valuable insights into the morphology of a wide variety of natural and synthetic sand-sized particles from different regions, with varying mineralogy and history. The primary goal of The Sand Atlas is to support researchers, educators, and industry professionals by providing detailed, easily accessible and uniformly produced surface meshes and level-set data. The underlying code that converts volumetric data to meshes is also available \textit{via} the \texttt{sand-atlas} python package. This platform encourages community participation, inviting contributors to share their own data and enrich the collective understanding of granular materials.
}


% \jnlcitation{\cname{%
% \author{Vego I.}, and
% \author{Marks B}}.
% \ctitle{On simplifying ‘incremental remap’-based transport schemes.} \cjournal{\it J Comput Phys.} \cvol{2021;00(00):1--18}.}

\maketitle

\renewcommand\thefootnote{}
\footnotetext{\textbf{Abbreviations:} DEM, discrete element method.} % separate with ;

\renewcommand\thefootnote{\fnsymbol{footnote}}
\setcounter{footnote}{1}

\section{Introduction}
Understanding the morphology and properties of sand-sized particles is essential in fields such as geotechnical engineering, sedimentology, and material science. The shape, size, and surface characteristics of sand-sized particles significantly influence the mechanical and flow behaviour of granular assemblies, affecting properties like strength, stiffness, and permeability \citep{santamarina2001soils, fonseca2012non, bofan2023shapeFlow}. Despite their importance, obtaining detailed morphological data of individual particles has been challenging due to limitations in imaging and processing technologies \citep{alert2022advanced}.

The Sand Atlas (\url{sand-atlas.scigem.com}) addresses this gap by collating a comprehensive dataset of sand-sized grains, with detailed 3D surface meshes generated from micro-computed tomography (micro-CT) scans. Micro-CT scanning allows for non-destructive, high-resolution imaging of the internal and external structures of particles \citep{ketcham2001acquisition}. Each dataset in The Sand Atlas is available at voxel-scale accuracy, as well as four standardised resolutions, enabling researchers to explore and download models in varying quality suitable for different applications.

This project is a collaborative effort between international researchers specialising in granular materials. The platform is designed to be a community-driven resource, encouraging contributions from the global scientific community. By providing open access to high-quality morphological data, The Sand Atlas aims to facilitate advancements in research areas such as particle shape analysis \citep{vlahinic2014towards}, discrete element method (DEM) simulations \citep{kawamoto2018all}, and soil mechanics \citep{altuhafi2016effect}.

\begin{table*}[ht]
\centering
\caption{Summary of available datasets on The Sand Atlas as of 06/02/2025.\label{tab:summary}}
\begin{tabular}{lllll}
\textbf{Sample Name} & \textbf{Type of Particle} & \textbf{Number of Particles} & \textbf{Voxel Size (\textmu m)} & \textbf{Source publication} \\ \hline
Couscous             & Edible grain              & 81                          & 2.8                              & \cite{vego2023thesis}\\
Hamburg sand         & Quartz quarry sand        & 8088                        & 11                               & \cite{milatz2021quantitative} \\
Hostun sand          & Angular quartz sand       & 86                          & 1                                & \cite{Wiebicke_2017}\\
Operculina ammonoides& Foraminifera species      & 46                          & 15                               & \cite{luijmes2024forametcetera}\\
Ottawa sand 7.3µm    & Natural silica sand       & 3367                        & 7.3                              & \\
Ottawa sand 1µm      & Natural silica sand       & 244                         & 1                                & \\
LECA                 & Expanded clay             & 598                         & 10.75                            & \cite{guida2018breakage} \\
Caicos Ooids         & Calcitic ooids            & 117                         & 1                                & \\
Coffee               & Edible grain              & 1                           & 0.8                              & \\
\end{tabular}
\end{table*}

The data processing pipeline uses the open source tools \texttt{numpy} \citep{harris2020array}, \texttt{scikit-image} \citep{scikit-image}, \texttt{spam} \citep{spam}, Blender \citep{blender} and OpenVDB \citep{museth2013vdb}, which ensure high-quality and reproducible data outputs. \texttt{numpy} is a python package for maniuplating arrays. \texttt{scikit-image} and \texttt{spam} are collections of algorithms for image processing. Blender is a 3D modelling program with advanced mesh operations. OpenVDB is an C++ library comprising a novel data structure and tools for the efficient storage and manipulation of sparse volumetric data. The underlying code that converts volumetric data to meshes is also available via the \texttt{sand-atlas} python package, promoting transparency and enabling users to process their own datasets according to their needs.

The 3D models available on the platform represent a wide range of particles, including geological samples such as Ottawa sand, which is a standardised testing sand in geotechnical engineering \citep{ASTM_C778}, and industrial materials such as Hostun sand, known for its angular grains \citep{Wiebicke_2017}. By encompassing a variety of particle types with different origins and properties, The Sand Atlas aims to provide a valuable resource for comparative studies and modelling efforts.

The broader goal of The Sand Atlas is to support the advancement of knowledge in the field of granular materials by providing a repository of high-resolution particle data. This data set has significant reuse potential, enabling applications in numerical simulations, educational tools, and the development of new analytical and statistical methods for particle characterisation.

\section{Data Description and Development}

\begin{figure*}[t]
    \centering
    \includegraphics[width=0.83333\textwidth]{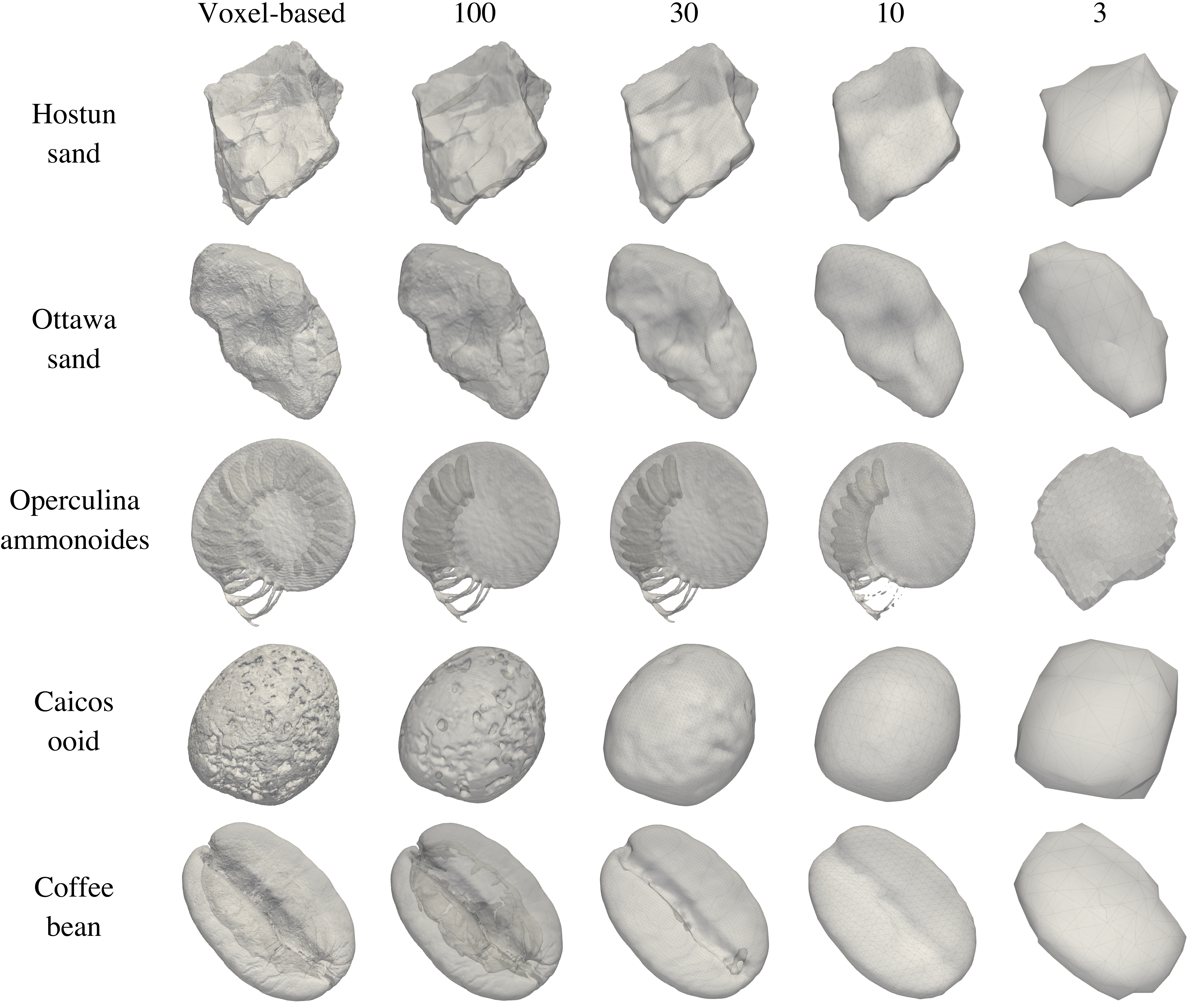}
    \caption{Examples of the meshing procedure applied to some particles from The Sand Atlas. From top to bottom: Hostun sand, Ottawa sand, Operculina ammonoides, Caicos ooids, and coffee bean \citep{Wiebicke_2017, luijmes2024forametcetera}. The left column illustrates meshes produced with a resolution equal to the voxel size. The remaining columns display meshes with resolutions of 100, 30, 10, and 3 elements along the shortest axis, respectively.}
    \label{fig:hostun}
\end{figure*}

\begin{figure*}
    \centering
    \includegraphics[width=0.83333\textwidth]{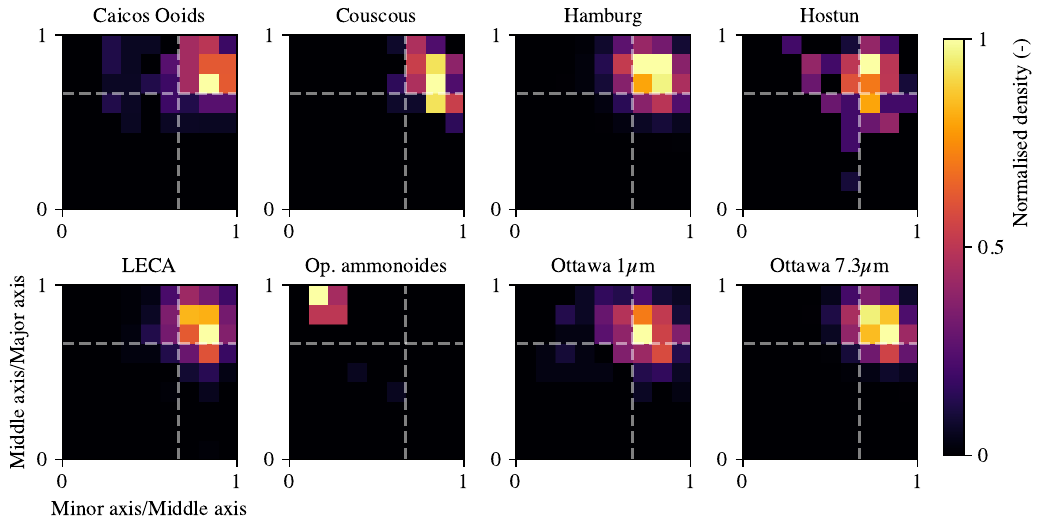}
    \caption{Zingg diagram \citep{zingg1935beitrag} representing particle shape statistics. Dashed white lines represent classification boundaries between regions of discoids, spheroids, spindles and ellipsoids (clockwise from top left).}
    \label{fig:zingg}
\end{figure*}

A snapshot of the available data is shown in Table~\ref{tab:summary}, encompassing 12,628 unique particles from nine different data sources. Examples of the rendered meshes of some of these particles are shown in Figure~\ref{fig:hostun} in various mesh qualities, each of which is available to download. Statistical measures of the shapes of the particles have also been computed for each particle, and representative measures for most samples are shown in Figure~\ref{fig:zingg}.

\subsection{Data pipeline}

\begin{figure*}
    \centering
    \includegraphics{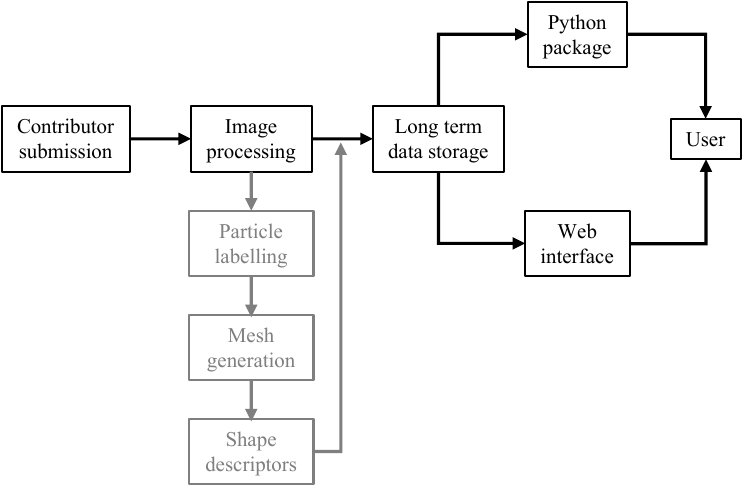}
    \caption{Schematic representation of The Sand Atlas data pipeline. The data flow from contributor submission to user download is represented in black, with options for download via the python package or web interface. The image processing pipeline is shown in grey.}
    \label{fig:schematic}
\end{figure*}

As outlined in Figure~\ref{fig:schematic}, the creation of The Sand Atlas dataset involves several key steps: sample collection, image processing, mesh generation, video rendering, and data dissemination. The first two steps are performed by contributors prior to uploading their data to the database, although the second step may be reproduced by The Sand Atlas Editorial team if the submitted data does not meet the quality thresholds.

\begin{enumerate}
    \item \textbf{Contributor submission:} Samples are submitted to the repository of particles from various natural and industrial sources, ensuring a diverse representation of particle types. Each sample is carefully documented with information -- when available -- on its origin, mineralogy, and any relevant geological or industrial processing history. Typically the source data has been uploaded to another data repository, and already has been assigned a DOI.

    \item \textbf{Image Processing and Segmentation:} The raw micro-CT data is processed to enhance image quality and segment individual particles. Noise reduction filters and thresholding techniques are applied to separate particles from the background. The segmentation process isolates each particle, creating labelled datasets for further analysis.

    \item \textbf{Mesh Generation:} Surface meshes are generated from labelled volumetric data using OpenVDB, allowing for efficient storage and manipulation of the sparse volumetric data, facilitating the conversion to high-quality surface meshes. The meshes are produced at five levels of resolution to accommodate different research needs.

    \item \textbf{Video generation:} Blender is used to render videos of each particle, so that a user can visualise and inspect the particles and the mesh quality on The Sand Atlas website before downloading.

    \item \textbf{Shape Descriptor Calculation:} Shape descriptors such as sphericity, roundness, and aspect ratio are calculated using \texttt{spam} \citep{spam}. These quantitative measures provide valuable information about the morphological characteristics of each particle.

    \item \textbf{Data Dissemination:} The processed data, including volumetric data (VDB files), surface meshes (STL files), and calculated shape descriptors, are uploaded to The Sand Atlas platform. Metadata accompanying each dataset includes information about the sample origin, scanning parameters, and processing techniques.
\end{enumerate}

\subsection{Technical Validation}

%This section presents any experiments or analyses that are needed to support the technical quality of the dataset. This section may be supported by figures and tables, as needed. This is a required section; authors must present information justifying the reliability of their data.

To ensure the reliability and accuracy of the dataset, several validation steps were performed. Upon submission, micro-CT images are inspected for artefacts, noise levels, and resolution adequacy. The segmentation results are validated by comparing the segmented particles with the original images to ensure accurate boundary detection. Over- and under-segmentation issues are resolved using the algorithms proposed and implemented in \texttt{spam} \citep{spam}. If necessary, the images are relabelled by The Sand Atlas team. Meshes are refined or regenerated if any issues are detected. The calculated shape descriptors are cross-validated with known standards or reference measurements where available.

\section{Dataset Access}

The Sand Atlas hosts multiple datasets, each corresponding to a different type of sample. These datasets are stored on the Research Data Store infrastructure provided by The University of Sydney and are accessible through The Sand Atlas website or python package. Each dataset is given a Unique Resource Identifier (URI) when first added to the repository. Each dataset includes:

\begin{enumerate}
    \item Volumetric Data: The original raw micro-CT image and the corresponding labelled image are stored as TIFF files.
    \item Level Set Representation: VDB files containing the level set representations of the particles.
    \item Surface Meshes: STL files of the particles at varying mesh resolutions.
    \item Shape Descriptors: CSV files containing calculated shape parameters for each particle (\textit{e.g.}, volume, equivalent diameter, axes lengths).
    \item Metadata: Documentation of the sample origin, scanning parameters, and processing methods.
\end{enumerate}

For example, the Ottawa 7.3\textmu m sand dataset contains 3,367 individual particles with mesh resolutions ranging from 1 facet per voxel to 100, 30, 10 and 3 facets across the shortest axis. Users have the flexibility to adjust the resolution and file formats to meet their research requirements by directly modifying the open-source code. These datasets can be used for applications such as DEM simulations, morphological analysis, and educational purposes.

\section{Potential Dataset Use and Reuse}

Each dataset in The Sand Atlas is freely available under a Creative Commons Attribution license, allowing users to share and adapt the material provided appropriate credit is given. Researchers can access and download the raw (\textit{i.e.}, directly provided by the contributor) and processed data directly from The Sand Atlas website. Users can also download the data using the \texttt{sand-atlas} python package.

To use the datasets:
\begin{itemize}
    \item Visualization: The STL mesh files can be viewed using standard 3D visualisation software such as MeshLab and Paraview.
    \item Simulation: The mesh and volumetric data can be imported into simulation software for DEM or finite element analyses.
    \item Analysis: Shape descriptors can be used for statistical analysis or to correlate particle morphology with material properties.
\end{itemize}

Contributors are encouraged to submit their own data through the submission process outlined on the website. The platform supports visual exploration, allowing users to manipulate 3D models directly on the website.

The Sand Atlas repository is expected to expand over time as new datasets are added. Future updates may include additional sand samples and granular materials contributed by the research community or generated through laboratory studies. These efforts aim to further enhance the diversity and applicability of the repository across various scientific disciplines, including but not limited to geoscience and geomechanics.

\subsection{Code availability}

% For all studies using custom code in the generation or processing of datasets, a statement must be included under the heading "Code availability", indicating whether and how the code can be accessed, including any restrictions to access. This section should also include information on the versions of any software used, if relevant, and any specific variables or parameters used to generate, test, or process the current dataset.
Two separate code sources are available. Firstly, the code used to generate the website itself is available at \url{https://github.com/scigem/sand-atlas/}. This includes the search functionality and the interactive elements.

The code used to process the micro-CT data and generate the meshes is available as the \texttt{sand-atlas} python package. The source code for this package is stored on GitHub at \url{https://github.com/scigem/sand-atlas-python/}. This package includes scripts for image processing, segmentation, mesh generation, and calculation of shape descriptors. Specific variables and parameters used in the processing pipeline are documented within the code repository, along with examples and instructions for reproducing the datasets.

\subsection{Data availability}

All data stored in The Sand Atlas is freely available under a Creative Commons licence. Data can be downloaded directly from \url{https://sand-atlas.scigem.com} or via the \texttt{sand-atlas} python package.

\bibliography{sample}

% \noindent LaTeX formats citations and references automatically using the bibliography records in your .bib file, which you can edit via the project menu. Use the cite command for an inline citation, e.g. \cite{Kaufman2020, Figueredo:2009dg, Babichev2002, behringer2014manipulating}. For data citations of datasets uploaded to e.g. \emph{figshare}, please use the \verb|howpublished| option in the bib entry to specify the platform and the link, as in the \verb|Hao:gidmaps:2014| example in the sample bibliography file. For journal articles, DOIs should be included for works in press that do not yet have volume or page numbers. For other journal articles, DOIs should be included uniformly for all articles or not at all. We recommend that you encode all DOIs in your bibtex database as full URLs, e.g. https://doi.org/10.1007/s12110-009-9068-2.

\section*{Acknowledgements}
    
The authors thank the contributing researchers and the development team for their ongoing efforts in maintaining and expanding The Sand Atlas. Special recognition goes to the core steering committee for their leadership in this project. The authors would also like to thank the organisers of the ``Getting into Shape'' Lorentz Centre Workshop where this idea was discussed and significantly progressed.

\section*{Author contributions statement}

B.M. conceived of and designed the project. B.M. developed the data processing pipeline and The Sand Atlas platform. I.V. performed the data processing. Both authors analysed the results and reviewed the manuscript.

\section*{Conflicts of interest}

The authors declare no conflict of interest.

\end{document}